\documentclass[sigconf]{acmart}
\usepackage{graphicx}
\usepackage{algorithm}
\usepackage{rotating}
\usepackage{balance}

\usepackage{tabularx}
\usepackage{subfigure}
\usepackage{bm}
\usepackage{multirow}
\usepackage{enumitem}
\AtBeginDocument{%
  \providecommand\BibTeX{{%
    \normalfont B\kern-0.5em{\scshape i\kern-0.25em b}\kern-0.8em\TeX}}}

\copyrightyear{2021}
\acmYear{2021}
\setcopyright{acmcopyright}\acmConference[KDD '21]{Proceedings of the 27th ACM SIGKDD Conference on Knowledge Discovery and Data Mining}{August 14--18, 2021}{Virtual Event, Singapore}
\acmBooktitle{Proceedings of the 27th ACM SIGKDD Conference on Knowledge Discovery and Data Mining (KDD '21), August 14--18, 2021, Virtual Event, Singapore}
\acmPrice{15.00}
\acmDOI{10.1145/3447548.3467093}
\acmISBN{978-1-4503-8332-5/21/08}
\begin{document}
\fancyhead{}

\title{Learning to Expand Audience via Meta Hybrid Experts and Critics for Recommendation and Advertising}


\author{Yongchun Zhu$^{1,2,3}$, Yudan Liu$^{3}$, Ruobing Xie$^{3}$, Fuzhen Zhuang$^{4,5,*}$, Xiaobo Hao$^{3}$, Kaikai Ge$^{3}$, \\Xu Zhang$^3$, Leyu Lin$^3$ and Juan Cao$^{1,2}$}
\affiliation{%
 \institution{$^1$Key Lab of Intelligent Information Processing of Chinese Academy of Sciences (CAS), Institute of Computing Technology, CAS, Beijing 100190, China $^2$University of Chinese Academy of Sciences, Beijing 100049, China\\
 $^3$WeChat Search Application Department, Tencent, China\\
 $^4$Institute of Artificial Intelligence, Beihang University, Beijing 100191, China\\
 $^5$SKLSDE, School of Computer Science, Beihang University, Beijing 100191, China\\
 \{zhuyongchun18s, caojuan\}@ict.ac.cn, \{danydliu, ruobingxie, rolyhao, kavinge, xuonezhang, goshawklin\}@tencent.com,zhuangfuzhen@buaa.edu.cn}\country{}} 
% \affiliation{%
% \institution{$^2$University of Chinese Academy of Sciences, Beijing 100049, China}\country{}} 
% \affiliation{%
% \institution{$^3$WeChat Search Application Department, Tencent, China }\country{}} 
% \affiliation{%
% \institution{$^4$Institute of Artificial Intelligence, Beihang University, Beijing 100191, China.}\country{}} 
% \affiliation{%
% \institution{$^5$SKLSDE, School of Computer Science, Beihang University, Beijing 100191, China.}\country{}} 
% \affiliation{\{zhuyongchun18s, caojuan\}@ict.ac.cn, \{danydliu, ruobingxie, rolyhao, kavinge, xuonezhang, goshawklin\}@tencent.com,zhuangfuzhen@buaa.edu.cn\country{}}
%\email{{zhuyongchun18s, caojuan}@ict.ac.cn,{danydliu, xrbsnowing, kavinge, xuonezhang, goshawklin}@tencent.com,zhuangfuzhen@buaa.edu.cn}
\thanks{*Fuzhen Zhuang is the corresponding author.}

\renewcommand{\shortauthors}{Y. Zhu et al.}

\begin{abstract}
  In recommender systems and advertising platforms, marketers always want to deliver products, contents, or advertisements to potential audiences over media channels such as display, video, or social. Given a set of audiences or customers (seed users), the audience expansion technique (look-alike modeling) is a promising solution to identify more potential audiences, who are similar to the seed users and likely to finish the business goal of the target campaign. However, look-alike modeling faces two challenges: (1) In practice, a company could run hundreds of marketing campaigns to promote various contents within completely different categories every day, e.g., sports, politics, society. Thus, it is difficult to utilize a common method to expand audiences for all campaigns. (2) The seed set of a certain campaign could only cover limited users. Therefore, a customized approach based on such a seed set is likely to be overfitting.
  
  In this paper, to address these challenges, we propose a novel two-stage framework named Meta Hybrid Experts and Critics (MetaHeac) which has been deployed in WeChat Look-alike System. In the offline stage, a general model which can capture the relationships among various tasks is trained from a meta-learning perspective on all existing campaign tasks. In the online stage, for a new campaign, a customized model is learned with the given seed set based on the general model. 
  According to both offline and online experiments, the proposed MetaHeac shows superior effectiveness for both content marketing campaigns in recommender systems and advertising campaigns in advertising platforms. Besides, MetaHeac has been successfully deployed in WeChat for the promotion of both contents and advertisements, leading to great improvement in the quality of marketing. The code has been available at \url{https://github.com/easezyc/MetaHeac}.
  
%   In this paper, to address these challenges, we propose a novel two-stage framework named Meta Hybrid Experts and Critics (MetaHeac) which has been deployed in WeChat Look-alike System. In the offline stage, MetaHeac trains a general model on existing campaigns. In the online stage, for a new campaign, a customized model is learned with the given seed set based on the general model. MetaHeac adopts two key ideas: (1) The general model is expected to learn the ability to expand audiences and adapt fast to new campaigns. Therefore, the general model is trained via a two-phase simulation from a meta-learning perspective, which can enhance the generalization ability. (2) The general model should learn more transferable knowledge from existing campaigns, and it is important to model the relationships among various tasks. Thus, MetaHeac employs hybrid experts to extract user representations and hybrid critics to give final scores. 
%   According to both offline and online experiments, the proposed MetaHeac shows superior effectiveness for both content marketing campaigns in recommender systems and advertising campaigns in advertising platforms. Besides, MetaHeac has been successfully deployed in "Top Stories" and "Subscriptions" of WeChat for promotion of both contents and advertisements, leading to great improvement in the quality of marketing. The code has been available at \url{https://github.com/easezyc/MetaHeac}.
\end{abstract}





\maketitle

\vspace{-0.3cm}
\section{Introduction}
With the booming development of the mobile internet, there are more and more mobile applications in the world, and a large number of users are active on these applications every day. In such a new mobile market with billions of users, it becomes crucial for marketers to effectively deliver contents~\cite{liu2019real},  advertisements~\cite{dewet2019finding} or products~\cite{zhuang2020hubble,liu2020two} to potential audiences by recommender systems or advertising platforms. To achieve the goals of various promotion activities, Internet companies conduct hundreds of marketing campaigns every day. One of the keys of running marketing campaigns is the audience expansion technique (look-alike modeling) which has been deployed in many online systems, e.g., Google~\cite{kanagal2013focused}, Linkedin~\cite{liu2016audience}, Pinterest~\cite{dewet2019finding}, Ant Financial~\cite{zhuang2020hubble}, WeChat~\cite{liu2019real}.

%Given a task of a certain campaign which consists of a set of audiences (seed users) and campaign features, the audience expansion technique aims to identify more potential audiences, who are similar to the seed users and likely to finish the business goal of the target campaign.

For a certain campaign, given a set of audiences (seed users), the audience expansion technique aims to identify more potential audiences who are similar to the seed users and likely to finish the business goal of the target campaign. High-quality expanded audiences will benefit the companies by increasing the conversion population while reducing the operating costs for their marketing campaigns~\cite{zhuang2020hubble}. A good look-alike technique can result in a great economic benefit, but it suffers from two significant challenges. (1) The tasks of various campaigns can cover diverse contents, e.g., sports, politics, society, and the promoted contents can vary from task to task. Thus, it is hard to exploit a common method for all campaign tasks. (2) A certain campaign gives a seed set that can only cover limited users. Based on this seed set, a customized approach for the campaign is likely to be overfitting. Especially, some marketing campaigns only have a few hundred seeds.

With designed common rules for all campaigns, previous rule-based approaches~\cite{mangalampalli2011feature,shen2015effective} match similar users with specific demographic tags (age, gender, geography) or interests that are targeted by marketers, which has unsatisfying performance. %ignores sophisticated details of user behaviors.
Using common pre-defined similarity function objectively such as Cosine or Jaccard for all campaigns, traditional similarity-based look-alike methods~\cite{liu2016audience,ma2016sub,ma2016score,doan2019adversarial} usually find look-alike users by directly comparing all possible pairs between seed users and available users. However, the performance of these methods largely depends on selecting the right features and similarity functions, and it is hard to achieve satisfying performance. Recently, training customized prediction models for each campaign, model-based look-alike approaches~\cite{qu2014systems,jiang2019comprehensive,liu2019real} achieve remarkable results.
However, these one-stage model-based look-alike methods~\cite{qu2014systems,jiang2019comprehensive,liu2019real} which directly train customized models from scratch for each campaign seriously suffer from the overfitting problem. 

%Only given the seeds of a certain campaign, the customized model could be overfitting on the given seeds. Especially, some marketing campaigns only have limited seeds, e.g., a few hundred users. Therefore, these customized models have extremely unsatisfying performance.

%In practice, the seed lists of different marketing campaigns vary drastically in size, from a few hundred users up to tens of millions of users. (1) For the campaigns with relatively sufficient seed users, training a customized model to converge from scratch is time-consuming. (2) For the campaigns with limited seed users, the seeds usually fail to cover the entire actual audiences. Thus, the customized model trained with the limited seeds is likely to be overfitting.

To tackle these challenges, some model-based methods~\cite{dewet2019finding,zhuang2020hubble,liu2020two} decompose the audience expansion tasks into two stages: offline stage and online stage. In the offline stage, they train a common embedding layer. In the online stage, based on the common embedding, customized models are trained for each campaign. However, these methods cannot solve the two challenges for three reasons. (1) The goal of look-alike modeling is to perform well for new campaign tasks. Nevertheless, the common embedding is just fitted on existing campaigns, and the generalization ability~\cite{wang2021generalizing} is not taken into consideration. (2) They directly train the common embedding layer over a large user-campaign graph ignoring the relationships among various tasks. Actually, the inherent conflicts from task differences can actually harm the predictions of some tasks, e.g., the seed users of a certain task can be negative samples of another task.  (3) They only pre-train the embedding layer without the deep network. Thus, for a certain campaign, the customized network could be still overfitting to the seed users.

%To address this drawback, a direct solution is to learn a pre-trained model that contains transferable knowledge across various marketing campaigns. However, the relationships of different marketing campaigns are complicated, i.e., the relationships among tasks could be positively related, negatively related, or unrelated. In detail, different marketing campaign tasks pay attention to different user seeds. The user seeds of a certain task can be similar to user seeds or negative samples of an another task. Thus, it is extremely hard to train a common binary prediction model shared by all campaigns. Some methods~\cite{dewet2019finding,zhuang2020hubble,liu2020two} decompose the audience expansion tasks into two stages: offline stage and online stage. However, all of them only learn pre-trained user representations (embeddings) across all tasks in the offline stage, and then train a customized deep model from scratch based on these user representations for each campaign. However, these methods suffer from two issues, 1) The parameters of the deep network could be still overfitting to the seed users of a certain task. 2) They directly train the embeddings over a large user-campaign graph ignoring the relationships among various tasks. 3) The goal of look-alike modeling is that performing well for new campaign tasks. However, the generalization ability is not taken into consideration.

%Since the deep model is still trained from scratch, it is obvious that these methods cannot solve the two challenges above.

To address these issues, we propose a novel two-stage audience expansion framework named Meta Hybrid Experts and Critics (MetaHeac). The framework of  MetaHeac follows a hybrid online-offline manner. The core of the offline stage is to train the general model with various existing marketing campaign tasks. In the online stage, when a new marketing campaign comes, a customized model which serves for audience expansion is learned by fine-tuning the pre-trained general model with the given seed users. MetaHeac mainly focuses on training a better general model and adopts two high-level key ideas: 1) The general model is expected to learn the ability to expand audiences, 2) The general model should learn transferable knowledge from various marketing campaigns.

Recall that the look-alike modeling consists of two phases, (1) understand the characteristics of seed users, (2) find the potential audiences who are similar to the seed users. Thus, in order to learn how to expand audiences, we design a two-phase simulation over hundreds of marketing campaign tasks. The simulation consists of an `understanding' phase and a `finding' phase. At the `understanding' phase, the model needs to understand the characteristics of the seed users. At the `finding' phase when we have found some potential audiences, we update the model with the expanded audiences. With the advantage of meta-learning~\cite{finn2017model}, the model trained on learning tasks of such two-phase simulation can learn how to expand audiences.

To learn transferable knowledge, it is important to model the relationships among tasks. Thus, MetaHeac takes a hybrid structure that consists of multiple experts and critics. MetaHeac utilizes multiple experts to capture multiple user representations. Since different experts specialize in different tasks, we propose a task-driven gate to combine user's multiple representations into a single representation. Then, with the user representation, MetaHeac employs multiple critics to express their opinions about whether the user can be interested in the promoted contents. Having the scores given by each critic, a similar task-driven gate is exploited to count the final score of the user.

MetaHeac has been successfully deployed in WeChat Look-alike System, and it can work for WeChat to promote both contents and advertisements, leading to great improvement in the quality of marketing. According to both offline and online experiments, the proposed model shows superior effectiveness for both marketing campaigns in recommender systems and advertising campaigns in advertising platforms.

The main contributions of our work are summarized into four folds:
\begin{itemize}
    \item We propose a novel two-stage framework named MetaHeac for the audience expansion problem, which significantly improves the performance.
    \item We formally define the audience expansion problem from a meta-learning perspective and use meta-learning to learn how to expand audiences.
    \item To learn transferable knowledge and model the task relationships, we propose a hybrid structure that consists of multiple experts and critics.
    \item We conduct both offline and online experiments to demonstrate the effectiveness and robustness of MetaHeac.
\end{itemize}
\section{System Overview}
\begin{figure}[t]
	\centering
	\begin{minipage}[b]{0.95\linewidth}
		\centering
		\includegraphics[width=1.\linewidth]{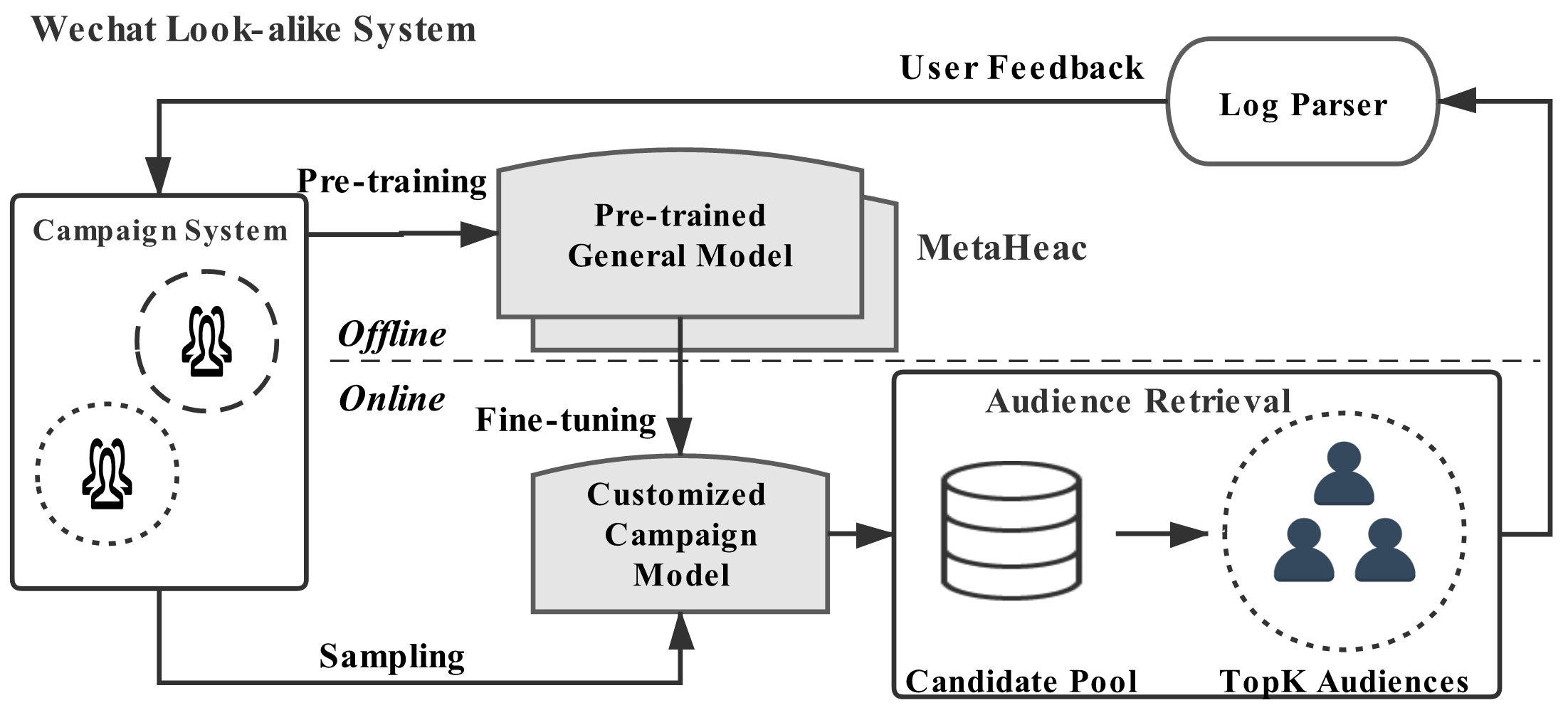}
	\end{minipage}
	\caption{The architecture of WeChat Look-alike System.}\label{fig:system}
\end{figure}

In this section, we will introduce the WeChat Look-alike System which contains the proposed MetaHeac framework. As shown in Figure~\ref{fig:system}, the system employs a hybrid online-offline architecture. Campaign System is a campaign management system that contains both seeds and online feedback of all campaigns. 

The core of the offline stage is to maintain a pre-trained general model that can adapt fast to new campaigns. Thus, WeChat Look-alike System utilizes data of all existing campaigns from the campaign system to train the general model. Note that the data consists of both the initial seeds and the online user feedback of all campaigns. To train the general model, the proposed MetaHeac framework is exploited. To learn how to expand audience, MetaHeac trains the general model from a meta-learning perspective. To capture task relationships and learn transferable knowledge, MetaHeac employs hybrid experts and critics. The details of MetaHeac will be introduced in the next section.

The online stage aims to find potential audiences for a certain campaign. For a new campaign, the marketers will give a certain seed set, and the campaign system will sample a negative set for training. With the data, a customized campaign model can be learned by fine-tuning the pre-trained general model. Then, with the customized model, the audience retrieval module will score all users in the candidate pool w.r.t. the target campaign, and then select the top K users with the highest scores as expanded audiences. Then, the contents of the target campaign will be promoted to the expanded audiences. The Log Parser will record the user feedback and pass it to the campaign system. The user feedback will be utilized to incrementally train the general model and the customized model.

\textbf{Remark.} The WeChat Look-alike System runs 400-500 marketing campaigns every day. The pre-trained general model is trained once a day with the new data of the day. The training of the customized model takes about 5 minutes. The audience retrieval of one billion users takes about 10 minutes. The customized model will be incrementally trained with real-time feedback once an hour.
\section{The Proposed Model}

\subsection{Definitions}
For a certain marketing campaign $c$, the marketers want to promote a set of contents or products denoted as $\mathcal{V}$. A marketing campaign has its specific goal, e.g., promoting a category of contents. Thus, the campaign has some features which can represent the commonalities of the promoted contents, e.g., categories, tags, creative size. The campaign $c$ can be denoted as $c = \{x^c_1,..., x^c_N\}$, where $x$ represents feature and $N$ is the number of features. Moreover, the marketers will give a set of seed users denoted as $\mathcal{S}_c$ which may be generated in different ways, e.g., selected by experts, retrieved according to the past relevant campaigns. To train a binary prediction model, both positive and negative samples are needed. Thus, the seed users $\mathcal{S}_{[c]}$ are labeled as positive, and the negative samples can come from random sampling on a candidate pool $\mathcal{U}$. And we denote the dataset consisting of both positive seeds and negative samples as $\mathcal{D}_{[c]}$. With $\mathcal{D}_{[c]}$ of a certain campaign, audience expansion aims to find a larger amount of users $\mathcal{U}_{[c]}$ (called expanded audiences) who are similar to users in $\mathcal{S}_{[c]}$ and likely to finish the business goal of the target campaign $c$. Both $\mathcal{D}_{[c]}$ and $\mathcal{U}_{[c]}$ come from the same candidate pool $\mathcal{U}$ denoted as $\mathcal{D}_{[c]},\mathcal{U}_{[c]} \subset \mathcal{U}$. Note that the size of $\mathcal{U}_{[c]}$ is usually given by the marketer of the campaign $c$ and is much larger than the size of $\mathcal{S}_{[c]}$. Each user $u \in \mathcal{U}$ may have multiple features, e.g., age, gender, city, interest tags. Therefore, a user can be denoted as $u = \{x^u_1,..., x^u_M\}$, where $M$ denotes the number of user's feature.

Then we need to encode the raw features $x$ into vectors. Look-up embedding has been widely adopted to learn dense representations from raw data for online prediction~\cite{pan2019warm,zhu2021learning}. Following~\cite{pan2019warm,zhu2020modeling}, we transform all features $x$ into embeddings denoted as $\bm{e}$. Then, we denote the dense representations of a campaign and a user as $\bm{c} = \{\bm{e}_1^c,...,\bm{e}^c_N\}$ and $\bm{u} = \{\bm{e}_1^u,...,\bm{e}^u_M\}$, where $\bm{e} \in \mathbb{R}^{k}$ and $k$ denotes the dimension of embeddings.

\begin{algorithm} [t] %\small
    \caption{Training MetaHeac from a meta-learning perspective.}\label{alg}
    \flushleft{\textbf{Input}: Given hundreds of marketing campaign dataset $\mathcal{D}_{[c]}$.
    
    \textbf{Input}: The general model $f_\theta$.
    
    \textbf{Input}: The learning rate $\alpha, \beta$.
    
    \begin{enumerate}[leftmargin=15pt]
        \item[1.] randomly initialize $\theta$.
        \item[2.] \textbf{while} not converge \textbf{do}:
        \item[3.] \qquad sample batch of training tasks \{$\mathcal{T}_1, ..., \mathcal{T}_n$\}.
        \item[4.] \qquad \textbf{for} $\mathcal{T}_i \in \{\mathcal{T}_1, ..., \mathcal{T}_n\}$ do:
        \item[5.] \qquad \qquad $\mathcal{T}_i$ contains two disjoint sets $\mathcal{D}^a_{[c]}, \mathcal{D}^b_{[c]}$
        \item[6.] \qquad \qquad evaluate loss $\mathcal{L}_a(\theta)$ with $\mathcal{D}^a_{[c]}$
        \item[7.] \qquad \qquad compute updated parameter $\theta_{[c]} = \theta - \alpha \frac{\partial \mathcal{L}_a(\theta)}{\partial \theta}$
        \item[8.] \qquad \qquad evaluate loss $\mathcal{L}_b({\theta_{[c]}})$ with $\mathcal{D}^b_{[c]}$
        \item[9.] \qquad \textbf{end}
        \item[10.] \qquad update $\theta = \theta - \beta \sum_{\mathcal{T}_i \in \{\mathcal{T}_1, ...,\mathcal{T}_n\}} \frac{\partial \mathcal{L}_b({\theta_{[c]}})}{\partial \theta}$
        \item[11.] \textbf{end}
    \end{enumerate}
    }
\end{algorithm}
%\vspace{-0.1cm}
\subsection{Learn to Expand Audience}\label{sec:meta}
A good look-alike model should be able to adapt to new marketing campaign tasks fast and alleviate the overfitting problem. Recently, meta-learning is a promising solution to achieve these goals: 1) meta-learning can learn general knowledge from lots of similar tasks~\cite{sun2019meta}, 2) meta-learning can achieve fast adaptation even only limited samples available~\cite{finn2017model}.
Thus, we propose a novel look-alike method from a meta-learning perspective. The key idea is learning how to expand audiences.

For a certain marketing campaign $c$, the final goal of audience expansion is to learn a customized model $f_{[c]}(\cdot;\theta_{[c]})$ which can find the potential audiences from the candidate pool $\mathcal{U}$. The input of the model $f_{[c]}(\cdot;\theta_{[c]})$ consists of both user and campaign features. Then, the binary prediction model can be denoted as:
\begin{equation}
    \hat{p} = f_{[c]} (\bm{c},\bm{u};\theta_{[c]}),
\end{equation}
where $\theta_{[c]}$ denotes the paramters of customized model $f_{[c]}(\cdot;\theta_{[c]})$.

Note that the procedure of audience expansion can be divided into two phases: 1) `Understanding': This phase aims to capture the characteristics of seed users $\mathcal{S}_{[c]}$ by training a customized model $f_{[c]}(\cdot)$ with the dataset $\mathcal{D}_{[c]}$. 2) `Finding': The `finding' phase exploits the customized model $f_{[c]}(\cdot)$ to seek an expanded audience set $\mathcal{U}_{[c]}$ from the candidate pool $\mathcal{U}$. To learn a general pre-trained model that knows how to expand audiences, we propose a training procedure to simulate the two phases. 

To begin with, we denote $f(\cdot;\theta)$ as the general model, where $\theta$ is the general parameters. Note that the general model $f(\cdot;\theta)$ is different from the customized model $f_{[c]}(\cdot;\theta_{[c]})$. With the dataset $\mathcal{D}_{[c]}$ of a certain campaign, we construct a training task $\mathcal{T}$ by sampling two disjoint mini-batches from $\mathcal{D}_{[c]}$, and we denoted the two mini-batches as $\mathcal{D}_{[c]}^a$ and $\mathcal{D}_{[c]}^b$ to simulate the `understanding' and `finding' phases, respectively. $\mathcal{D}_{[c]}^a$ can be seen as the seed users and the randomly sampled negative examples, while $\mathcal{D}_{[c]}^b$ can be seen as the expanded audiences. The $\mathcal{D}_{[c]}^a$ and $\mathcal{D}_{[c]}^b$ are also called as the support set and the query set in the meta-learning literature.

\textbf{Understanding phase.} The meta-learner adapts the general $\theta$ to customized task-specific parameters $\theta_{[c]}$ w.r.t. the loss on the support set $\mathcal{D}_{[c]}^a$. Firstly, we make predictions using the general model $f(\cdot;\theta)$ on the support set $\mathcal{D}_{[c]}^a$ and get the predicted $\hat{p} = f(\bm{c},\bm{u};\theta)$. Then, the loss of this phase can be defined as:
\begin{equation}
    \mathcal{L}_a(\theta) = \sum_{\mathcal{D}_{[c]}^a} [-y \log \hat{p} - (1 - y) \log (1-\hat{p})],\label{eq:local}
\end{equation}
where $y$ is the label of the sample and $\mathcal{L}_a(\theta)$ denotes the loss computed based on $\theta$. By minimizing the loss $\mathcal{L}_a(\theta)$, a customized model $f_{[c]}(\cdot;\theta_{[c]})$ can be learned, where $\theta_{[c]} = \theta - \alpha \frac{\partial \mathcal{L}_a}{\partial \theta}$. The $\alpha$ denotes the learning rate of this phase.

\textbf{Finding phase.} After the understanding phase, a customized model $f_{[c]}(\cdot;\theta_{[c]})$ for the campaign $c$ is available. Similarly, based on $\theta_{[c]}$, we can compute the loss on the query set $\mathcal{D}_{[c]}^b$ as $\mathcal{L}_b(\theta_{[c]})$. In order to learn how to expand audience, we utilize the loss computed with the `expanded' audience set $\mathcal{D}_{[c]}^b$ on parameters $\theta_{[c]}$ to update the general parameters $\theta$, as:
\begin{equation}
    \theta = \theta - \beta \frac{\partial \mathcal{L}_b(\theta_{[c]})}{\partial \theta} = \theta - \beta \frac{\partial \mathcal{L}_b(\theta_{[c]})}{\partial \theta_{[c]}} \frac{\partial \theta_{[c]}}{\partial \theta}.\label{eq:global}
\end{equation}
% where
% \begin{equation}
%     \frac{\partial \theta_{[c]}}{\partial \theta} = 1 - \alpha \frac{\partial^2 \mathcal{L}_a}{\partial \theta}.
%     \vspace{0.1cm}
% \end{equation}
The $\beta$ denotes the learning rate of the finding phase. Note that the meta-optimization is performed over the model parameters $\theta$, whereas the objective $\mathcal{L}_b(\theta_{[c]})$ is computed using the updated model parameters $\theta_{[c]}$. 

\textbf{Training procedure.} For each iteration, we sample a batch of training tasks from different marketing campaigns. Then, we update the general model $f(\cdot;\theta)$ with all training tasks together with Equation~(\ref{eq:global}). It is obvious that the setting of the training tasks is similar to the audience expansion task, and it only updates the general model with the expanded audiences $\mathcal{D}_{[c]}^b$ after understanding the seed users $\mathcal{D}_{[c]}^a$. Thus, the general model is able to learn how to expand audiences. We come to the overall training algorithm from a meta-learning perspective, see Algorithm~\ref{alg}.

\subsection{Hybrid Experts}\label{sec:he}%\vspace{0.1cm}
To select potential audiences, it is necessary to generate high-quality user representations that can represent users’ intentions toward different campaigns. Intuitively, we can employ an expert (a feed-forward network) to extract the users' representations for all campaigns. However, it faces a serious problem with a single expert. Usually, an expert only specializes in one area. The user's representation extracted by the single expert could only contain partial information about the user, which cannot completely cover the characteristics of the user. As a result, while other characteristics of the user have been not observed by the expert, such a user representation is only good for a part of campaigns. To address the challenge, it is intuitive to employ multiple experts. Formally, we average the user's representation extracted by all experts, as:
\begin{equation}
    \bm{r} = \frac{1}{n} \sum_{i = 1}^{n} h_i(\bm{u}),
    %\vspace{-0.1cm}
\end{equation}
where $h_i(\cdot)$ denotes the $i$-th expert, and $n$ denotes the number of experts. However, the operation of average can remove the task-specific information, so the user representation $\bm{r}$ may not be good for the specific marketing campaign tasks. Note that different experts specialize in different areas, and they are good at handling different tasks. Therefore, for a certain marketing campaign, we want to select experts who specialize in the specific task. 

\begin{figure}[t]
	\centering
	\begin{minipage}[b]{0.95\linewidth}
		\centering
		\includegraphics[width=1.\linewidth]{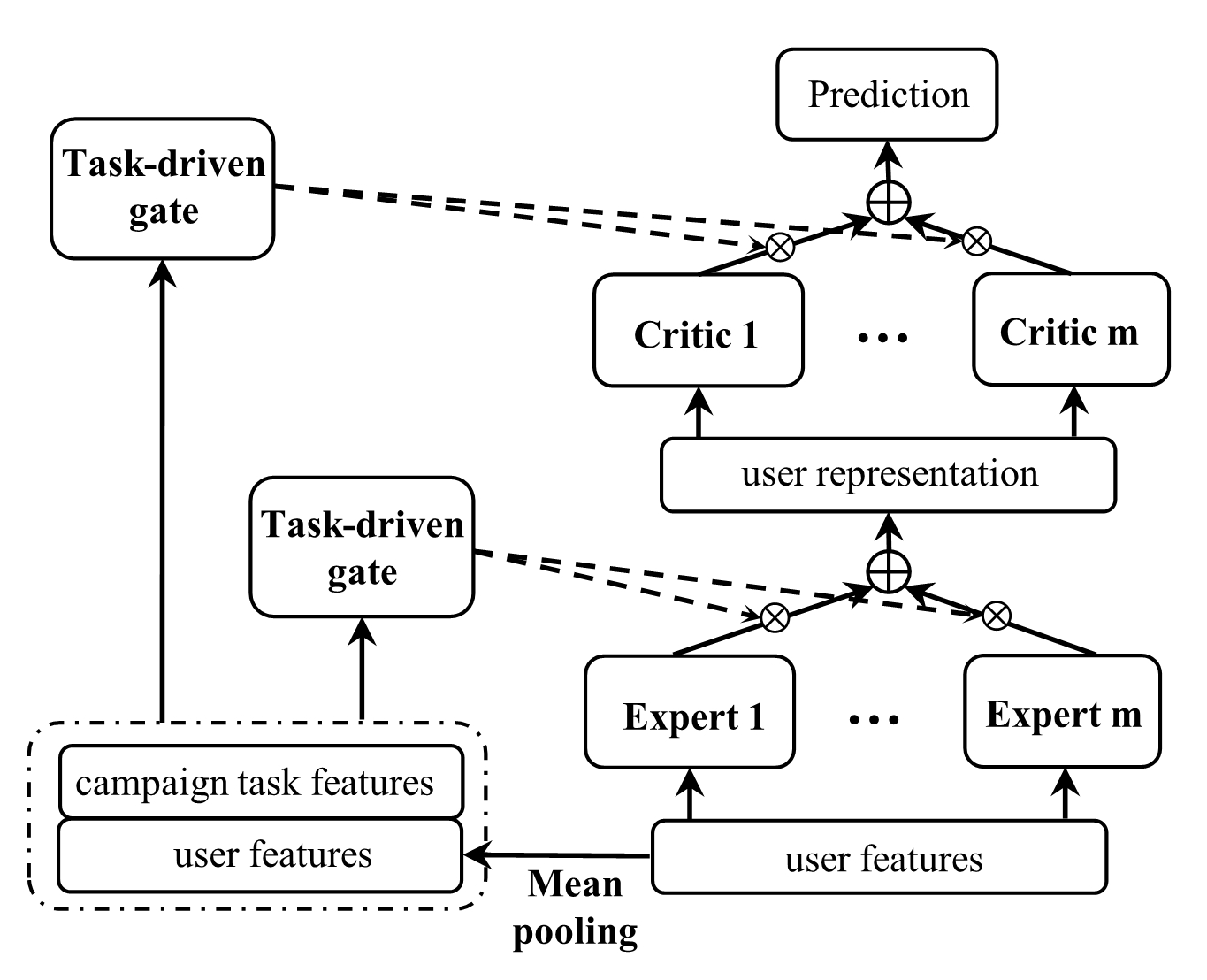}
	\end{minipage}%\vspace{-0.1cm}
	\caption{The model structure of the Meta Hybrid Experts and Critics (MetaHeac).}\label{fig:model}
\end{figure}

Along this line, we propose a task-driven gate with campaign features as input to guide the selection process. In addition, note that an expert only specializes in a part of tasks, and his ability to handle different users is also different. Thus, the task-driven gate should take not only campaign features as input but also the user features.
Thus, the task-driven gate is defined as:
\begin{equation}
%\vspace{-0.1cm}
    \bm{w}^{expert} = \text{softmax} (g(\bm{c}, \mathcal{G}(\bm{u}))),
    %\vspace{-0.05cm}
\end{equation}
where $g(\cdot)$ is a feed-forward network, and $\bm{c}$ and $\bm{u}$ are the feature embeddings of the campaign task $c$ and the user $u$. Besides, softmax denotes the softmax function to normalize the output of $g(\cdot)$, and $\bm{w}^{expert} \in \mathbb{R}^{n}$ is the weight vector which can denote the importance of different experts. $\mathcal{G}(\cdot)$ denotes an aggregating function. In this paper, we utilize mean pooling as $\mathcal{G}(\cdot)$. The input of $g(\cdot)$ is the embeddings of campaign features and the mean embedding of user features. Note that the mean pooling operation on user features is extremely important. The main reason is that we mainly focus on capturing the transferable knowledge across various tasks, not users. Thus, such a mean pooling operation can not only make the task information dominate the gate, but also exploit the user information. With the task-driven gate, we can reformulate the user's representation as:
\begin{equation}
    %\vspace{-0.1cm}
    \bm{r} = \frac{1}{n} \sum_{i = 1}^{n} w^{expert}_i h_i(\bm{u}),\label{eq:w}
    %\vspace{-0.1cm}
\end{equation}
where $w^{expert}_i$ denotes the $i$-th element in the vector $\bm{w}^{expert}$.

Each expert can specialize in different tasks, and the task-driven gate can select a subset of experts who are good at handling the specific task. This is desirable for a flexible parameter sharing in the audience expansion problem which faces hundreds of marketing campaign tasks. It is easy to understand the hybrid experts are able to model the task relationships. If the tasks are highly related, the task-driven gate will select similar experts to extract user representations. If the tasks are less related, the gate networks will learn to utilize different experts instead.

%\vspace{-0.1cm}
\subsection{Hybrid Critics}\label{sec:hc}
Given the user representation $\bm{r}$ extracted by hybrid experts, we need to decide whether this user is interested in the contents promoted by the certain campaign. There are two direct ways: 1) a single critic; 2) for each task, employ a critic. However, both of the two methods cannot achieve satisfying performance. For the first method, since a user could have different labels in different tasks, a single critic is hard to satisfy all tasks. For the second method, since some tasks only have limited seed users, it is difficult to train a specific critic for such tasks. Based on the intuition that different critics are good at different tasks, we employ multiple critics. And a single critic is defined as:
\begin{equation}
    \hat{p} = \sigma (t(\bm{r})),
    %\vspace{-0.1cm}
\end{equation}
where $\sigma$ denotes the sigmoid function, and $t(\cdot)$ is a feed-forward network. Besides, $\hat{p}$ is the probability predicted by the critic. Based on the similar motivation as hybrid experts, we utilize another task-driven gate to guide the combination of multiple critics:
\begin{equation}
    \hat{p} = \frac{1}{m} \sum_{i=1}^{m} w^{critic}_i \sigma (t_i(\bm{r})),
    \vspace{-0.1cm}
\end{equation}
where $t_i(\cdot)$ denotes the $i$-th critic, and $m$ is the number of critics. $w_i^{critic}$ denotes the weight of the $i$-th critic, which is similar to $w^{expert}_i$ in Equation~(\ref{eq:w}). Note that the task-driven gates for hybrid experts and hybrid critics are two separate modules.

It is obvious that the hybrid critics can also learn the relationships among various tasks as hybrid experts do.

\vspace{-0.1cm}
\subsection{Overall Framework}
The overall MetaHeac framework consists of the meta-learning framework, hybrid experts, and hybrid critics, as shown in Figure~\ref{fig:model}. Training the general look-alike model $f(\cdot)$ with the meta-learning framework, the model can learn how to expand audiences. In this paper, the model $f(\cdot)$ contains both hybrid experts and hybrid critics, formulated as:
\begin{equation}
    f(\bm{c},\bm{u}) = \frac{1}{m} \sum_{i=1}^{m} w^{critic}_i \sigma (t_i(  \frac{1}{n} \sum_{i = 1}^{n} w^{expert}_i h_i(\bm{u})  )).
\end{equation}
With such hybrid experts and critics, the look-alike model can capture the relationships among various tasks and learn transferable knowledge. The deployment of the MetaHeac framework consists of two stages: offline and online.

\textbf{Offline stage.} In this stage, we learn the general look-alike model with the meta-learning framework of the two-phases simulation on all existing marketing campaigns. All existing campaign tasks contain a set of seed users and the online feedback, and we combine all of them into a single training set to train the general model as Section~\ref{sec:meta}.

\textbf{Online stage.} For a new marketing campaign $c$, given the dataset $\mathcal{D}_{[c]}$ of the new campaign, we fine-tune the general look-alike model $f(\cdot;\theta)$ with the cross-entropy loss similar to Equation~(\ref{eq:local}). And we can obtain a customized look-alike model $f_{[c]}(\cdot;\theta_{[c]})$. Then, the customized model can be directly exploited to find the potential audiences $\mathcal{U}_{[c]}$ from the whole candidate pool $\mathcal{U}$ who could be interested in the contents promoted by the specific campaign $c$. Finally, the look-alike systems can promote the contents to the expanded users. 
\begin{table}[t]
  \centering
  \caption{Statistics of the experimental datasets. The `tasks' denotes the number of marketing campaigns.}
    \begin{tabular}{cccc}
    \toprule
          & tasks & seeds & expanded audiences \\
    \midrule
    Tencent Dataset & 172   & 421,961 & 2,265,989 \\
    WeChat Dataset & 90    & 286,036 & 3,146,396 \\
    \bottomrule
    \end{tabular}%
  \label{tab:dataset}%
\end{table}%

\section{Experiments}
In this section, we conduct extensive offline and online experiments with the aim of answering the following research questions: \textbf{RQ1:} Does our proposed MetaHeac outperform other look-alike approaches in different tasks?
\textbf{RQ2:} Does this MetaHeac framework get improvement on the performance of WeChat Look-alike system?
\textbf{RQ3:} What are the effects of meta-learning, hybrid experts, and hybrid critics in our proposed MetaHeac?
    %\item[\textbf{RQ4}] How does the MetaHeac capture the relationships among various audience expansion tasks?

\begin{table*}[t]
  \centering
  \caption{Offline results (AUC, P@5\% and R@5\%) on two look-alike datasets. We report the mean results over five runs. Best results are in boldface, and the second bests results are with underline. $**$ and $*$ indicates $0.005$ and $0.05$, paired t-test of MetaHeac vs. the best baselines. $Improve$ denotes relative improvement over the best baseline.}
    \begin{tabular}{c||c||cc||ccc||ccc}
    \toprule
    \multirow{2}[2]{*}{Dataset} & \multirow{2}[2]{*}{Method} & \multicolumn{2}{c||}{Pre-trained} & \multicolumn{3}{c||}{$\mathcal{S}_{[c]} \leq T$} & \multicolumn{3}{c}{$\mathcal{S}_{[c]} > T$} \\
\cline{3-10}          &       & \multicolumn{1}{c}{Emb} & \multicolumn{1}{c||}{Network} & \multicolumn{1}{c}{AUC} & \multicolumn{1}{c}{P@5\%} & \multicolumn{1}{c||}{R@5\%} & \multicolumn{1}{c}{AUC} & \multicolumn{1}{c}{P@5\%} & \multicolumn{1}{c}{R@5\%} \\
    \cline{1-10}
    \multirow{10}[5]{*}{\shortstack{Tencent\\Look-alike\\Dataset}} & LR    &   -    &    -   & \multicolumn{1}{c}{0.5942 } & \multicolumn{1}{c}{0.1015 } & \multicolumn{1}{c||}{0.1044 } & \multicolumn{1}{c}{0.6824 } & \multicolumn{1}{c}{0.1910 } & \multicolumn{1}{c}{0.2006 } \\
          & MLP\_one-stage &   -    &    -   & \multicolumn{1}{c}{0.5928 } & \multicolumn{1}{c}{0.1048 } & \multicolumn{1}{c||}{0.1081 } & \multicolumn{1}{c}{0.6910 } & \multicolumn{1}{c}{0.1797 } & \multicolumn{1}{c}{0.1888 } \\
\cline{2-10}          & MLP+emb &    $\checkmark$   & \multicolumn{1}{c||}{-} & \multicolumn{1}{c}{0.6624 } & \multicolumn{1}{c}{0.1881 } & \multicolumn{1}{c||}{0.1930 } & \multicolumn{1}{c}{0.7060 } & \multicolumn{1}{c}{0.2118 } & \multicolumn{1}{c}{0.2224 } \\
          & Pinterest &   $\checkmark$    & \multicolumn{1}{c||}{-} & \multicolumn{1}{c}{0.6245 } & \multicolumn{1}{c}{0.1635 } & \multicolumn{1}{c||}{0.1665 } & \multicolumn{1}{c}{0.6802 } & \multicolumn{1}{c}{0.1687 } & \multicolumn{1}{c}{0.1770 } \\
          & Hubble &   $\checkmark$    & \multicolumn{1}{c||}{-} & \multicolumn{1}{c}{0.6797 } & \multicolumn{1}{c}{0.2056 } & \multicolumn{1}{c||}{0.2110 } & \multicolumn{1}{c}{0.7085 } & \multicolumn{1}{c}{0.2171 } & \multicolumn{1}{c}{0.2279 } \\
\cline{2-10}          & MLP+pre-training & \multicolumn{1}{c}{$\checkmark$} & \multicolumn{1}{c||}{$\checkmark$} & \multicolumn{1}{c}{\underline{0.7117} } & \multicolumn{1}{c}{\underline{0.2325} } & \multicolumn{1}{c||}{\underline{0.2384} } & \multicolumn{1}{c}{0.7082 } & \multicolumn{1}{c}{0.2136 } & \multicolumn{1}{c}{0.2242 } \\
          & Shared-Bottom+pre-training & \multicolumn{1}{c}{$\checkmark$} & \multicolumn{1}{c||}{$\checkmark$} & \multicolumn{1}{c}{0.6936 } & \multicolumn{1}{c}{0.2198 } & \multicolumn{1}{c||}{0.2258 } & \multicolumn{1}{c}{\underline{0.7089} } & \multicolumn{1}{c}{0.2144 } & \multicolumn{1}{c}{0.2250 } \\
          & MMoE+pre-training  & \multicolumn{1}{c}{$\checkmark$} & \multicolumn{1}{c||}{$\checkmark$} & \multicolumn{1}{c}{0.6977 } & \multicolumn{1}{c}{0.2224 } & \multicolumn{1}{c||}{0.2280 } & \multicolumn{1}{c}{0.7088 } & \multicolumn{1}{c}{\underline{0.2150} } & \multicolumn{1}{c}{\underline{0.2257} } \\
%\cline{2-10}          
& MetaHeac & \multicolumn{1}{c}{$\checkmark$} & \multicolumn{1}{c||}{$\checkmark$} & \multicolumn{1}{c}{\textbf{0.7239** }} & \multicolumn{1}{c}{\textbf{0.2489** }} & \multicolumn{1}{c||}{\textbf{0.2554** }} & \multicolumn{1}{c}{\textbf{0.7142** }} & \multicolumn{1}{c}{\textbf{0.2244** }} & \multicolumn{1}{c}{\textbf{0.2356** }} \\
\cline{2-10}          & Improve &       &       & 1.7\% & 7.0\% & 7.1\% & 0.8\% & 4.7\% & 4.7\% \\
    \cline{1-10}
    \multirow{10}[5]{*}{\shortstack{WeChat\\Look-alike\\Dataset}} & LR    &   -    &    -   & \multicolumn{1}{c}{0.5654 } & \multicolumn{1}{c}{0.1351 } & \multicolumn{1}{c||}{0.0742 } & \multicolumn{1}{c}{0.6711 } & \multicolumn{1}{c}{0.2166 } & \multicolumn{1}{c}{0.1182 } \\
          & MLP\_one-stage &    -   &    -   & \multicolumn{1}{c}{0.6663 } & \multicolumn{1}{c}{0.2477 } & \multicolumn{1}{c||}{0.1363 } & \multicolumn{1}{c}{0.6970 } & \multicolumn{1}{c}{0.2605 } & \multicolumn{1}{c}{0.1419 } \\
\cline{2-10}          & MLP+emb &    $\checkmark$   & \multicolumn{1}{c||}{-} & \multicolumn{1}{c}{0.7143 } & \multicolumn{1}{c}{0.3058 } & \multicolumn{1}{c||}{0.1684 } & \multicolumn{1}{c}{0.7217 } & \multicolumn{1}{c}{0.2988 } & \multicolumn{1}{c}{0.1628 } \\
          & Pinterest &   $\checkmark$    & \multicolumn{1}{c||}{-} & \multicolumn{1}{c}{0.6289 } & \multicolumn{1}{c}{0.1947 } & \multicolumn{1}{c||}{0.1066 } & \multicolumn{1}{c}{0.7044 } & \multicolumn{1}{c}{0.2639 } & \multicolumn{1}{c}{0.1439 } \\
          & Hubble &    $\checkmark$   & \multicolumn{1}{c||}{-} & \multicolumn{1}{c}{0.7391 } & \multicolumn{1}{c}{\underline{0.3524} } & \multicolumn{1}{c||}{\underline{0.1936} } & \multicolumn{1}{c}{0.7243 } & \multicolumn{1}{c}{\underline{0.3062} } & \multicolumn{1}{c}{0.1668 } \\
\cline{2-10}          & MLP+pre-training & \multicolumn{1}{c}{$\checkmark$} & \multicolumn{1}{c||}{$\checkmark$} & \multicolumn{1}{c}{\underline{0.7440} } & \multicolumn{1}{c}{0.3473 } & \multicolumn{1}{c||}{0.1908 } & \multicolumn{1}{c}{0.7272 } & \multicolumn{1}{c}{0.3030 } & \multicolumn{1}{c}{0.1673 } \\
          & Shared-Bottom+pre-training & \multicolumn{1}{c}{$\checkmark$} & \multicolumn{1}{c||}{$\checkmark$} & \multicolumn{1}{c}{0.7271 } & \multicolumn{1}{c}{0.3093 } & \multicolumn{1}{c||}{0.1700 } & \multicolumn{1}{c}{0.7275 } & \multicolumn{1}{c}{0.3052 } & \multicolumn{1}{c}{0.1663 } \\
          & MMoE+pre-training  & \multicolumn{1}{c}{$\checkmark$} & \multicolumn{1}{c||}{$\checkmark$} & \multicolumn{1}{c}{0.7368 } & \multicolumn{1}{c}{0.3265 } & \multicolumn{1}{c||}{0.1797 } & \multicolumn{1}{c}{\underline{0.7292} } & \multicolumn{1}{c}{0.3051 } & \multicolumn{1}{c}{\underline{0.1675} } \\
%\cline{2-10}          
& MetaHeac & \multicolumn{1}{c}{$\checkmark$} & \multicolumn{1}{c||}{$\checkmark$} & \multicolumn{1}{c}{\textbf{0.7607**}} & \multicolumn{1}{c}{\textbf{0.3839**}} & \multicolumn{1}{c||}{\textbf{0.2110**}} & \multicolumn{1}{c}{\textbf{0.7323*}} & \multicolumn{1}{c}{\textbf{0.3133*}} & \multicolumn{1}{c}{\textbf{0.1707*}} \\
\cline{2-10}          & Improve &       &       &    2.3\%   &    8.9\%   &   9.0\%    &   0.4\%    &    2.3\%   &  1.9\%  \\
    \bottomrule
    \end{tabular}%
  \label{tab:offline}%
\end{table*}%

\subsection{Experimental Settings}\label{sec:exp}
\textbf{Datasets.} We evaluate MetaHeac with baselines on both large-scale public and private datasets of audience expansion tasks. 

\textit{Tencent Look-alike Dataset\footnote{https://algo.qq.com/archive.html?}:} The public dataset for Tencent Ads competitions in 2018 contains hundreds of seed sets and aims to find similar users. Each advertising task contains lots of seed users and non-seed users. Each advertisement has six categorical features: advertiser ID, ad category, campaign ID, product ID, product type,  creative size. Each user has 14 features, including carrier, consumption ability, location, age, education, gender, house, three-level interests, two-level keywords, two-level topics.

\textit{WeChat Look-alike Dataset\footnote{All data are preprocessed via data masking to protect the user's privacy.}:} The private dataset is from WeChat. This is a dataset of content promotion in recommender systems. Each task is a content marketing campaign that consists of several contents (items), lots of seed users, and non-seed users. We take the first few days to train the general model and test the performance on the tasks of the rest.

%All the campaigns occur in three days in 2021. We use the tasks of the first two days to train the general model and test the performance on the tasks of the third day. %Each user has seven features, including gender, age, city, elite level, three-level interests. Each content marketing campaign has four features: three-level categories, tags.

The details of the two datasets are shown in Table~\ref{tab:dataset}. For the Tencent and WeChat datasets, seed:non-seed = positive samples:  negative samples = 1:20 and 1:10, respectively. 
%In other words, in the training stage, positive samples:  negative samples = seeds : non-seeds. 
Each campaign $c$ contains a seed set $\mathcal{S}_{[c]}$ and a non-seed set for training, and a set of expanded audiences for testing. The set of expanded audiences consists of actual audiences (positive samples) and other candidate users (negative samples).

For WeChat Look-alike Dataset, we use all data of the first two days (58 tasks) to learn the pre-trained model, and the last day (32 tasks) for the test stage. Moreover, to validate the performance on campaigns with different size, according to the size of seed lists, we divide the campaigns of the test set into two groups: `$\mathcal{S}_{[c]} \leq T$' and `$\mathcal{S}_{[c]} > T$'. We set $T=10,000$ for WeChat Dataset. For the Tencent Look-alike Dataset, we randomly select 60\% campaigns as the training tasks, and others as the test tasks. We set $T=4,000$ for Tencent Dataset. About the selection of $T$, we keep the campaign tasks of `$\mathcal{S}_{[c]} \leq T$' : `$\mathcal{S}_{[c]} > T$' = 8:2.

\textbf{Baselines.} We categorize our baselines into three groups, and the first group is the traditional one-stage look-alike methods which directly train a model for the specific campaign without pre-training. (1) LR~\cite{qu2014systems}: a one-stage model-based look-alike method taking LR as the the classifier.
(2) MLP\_one-stage: an one-stage model-based method using MLP as the classifier. 

The second group is the two-stage approaches which only pre-train the embeddings.
(1) MLP+emb: all tasks share the same embedding layer, and a customized MLP is utilized for classification.
(2) Pinterest~\cite{dewet2019finding}: a two-stage approach presented by Pinterest. In the offline stage, a global embedding model is trained across all tasks. In the online stage, a similarity-based classification is performed on the pre-trained embeddings.
(3) Hubble~\cite{zhuang2020hubble}: a GNN-based two-stage method proposed by Ant Financial. In the offline stage, a GNN is trained on the large campaign-user graph to learn the user representations. In the online stage, based on the user representations, a customized MLP is learned for a certain campaign.

The third group is the two-stage methods that pre-train not only the embeddings but also the networks.
(1) MLP+pre-training: learn a general MLP (including embeddings) across various tasks in the offline stage, and fine-tune it for a specific campaign in the online stage.
(2) Shared-Bottom+pre-training~\cite{caruana1997multitask}: Shared-Bottom is a traditional multi-task model. We modify it to fit the audience expansion problem. The bottom shared by all tasks is pre-trained in the offline stage, and fine-tune the classification layer in the online stage.
(3) MMoE+pre-training~\cite{ma2018modeling}: MMoE is a recent popular multi-task model that shares a mixture-of-experts (MoE) across various tasks, and each task has its specific head. We pre-train the overall structure in the offline stage, and fine-tune it in the online stage for the new campaign.

% Actually, we also try to replace MLP with other advanced models, e.g., transformer~\cite{vaswani2017attention}, NFM~\cite{he2017neural}, AFN~\cite{cheng2020adaptive}. However, we find that these models cannot outperform MLP and is more time-consuming.

\textbf{Experimental details.} For all methods, the initial learning rate for the Adam~\cite{kingma2014adam} optimizer are tuned by grid searches within \{0.001, 0.005, 0.01, 0.02, 0.1\}, including the global learning rate $\beta$ in MetaHeac. Following~\cite{lu2020meta,zhu2021transfer}, we set the local learning rate $\alpha$ as 0.001. In addition, we set the dimension of embeddings as 16. For all methods, we set mini-batch size of 512. We tune the number of experts and critics in [3, 10]. For all MLP in these methods, the ReLU activation function is employed, and the dimension of each layer is set to 64, 64, and 1. For Shared-Bottom, MMoE, and MetaHeac, for a fair comparison, the experts and bottom are two-layer MLP, and the dimension of each layer is 64, 64. Besides, the heads or classifiers of the three methods are one fully-connected layer that outputs one unit. The gate network of MMoE and MetaHeac are two-layer MLP with dimension of each layer 64 and 1. For the two-stage methods, we train the model with the training set for one epoch in the offline stage, and we fine-tune the pre-trained model on $\mathcal{D}_{[c]}$ until loss converges in the online stage. For the one-stage methods, directly train the model for each campaign until loss converges.

\textbf{Evaluation metrics.} In binary prediction tasks, AUC (Area Under ROC) is  a widely used  metric~\cite{fawcett2006introduction} (to predict whether a user is actual audience or whether a conversion interaction exists between a user and a campaign, respectively). In addition, following~\cite{zhuang2020hubble,liu2020two}, we also choose another two metrics, Precision/Recall at top K percent candidates (P@K\% and R@K\% for short), defined as:
\begin{equation}\vspace{-0.1cm}
    \text{P@K\%} = \frac{|\mathcal{U}_{at} \cap \mathcal{U}_{cdd,K}|}{|\mathcal{U}_{cdd,K}|}, \quad \text{R@K\%} = \frac{|\mathcal{U}_{at} \cap \mathcal{U}_{cdd,K}|}{|\mathcal{U}_{at}|}.\vspace{-0.1cm}
\end{equation}
$\mathcal{U}_{at}$ denotes the set of actual audiences of the certain campaign, and $\mathcal{U}_{cdd,K}$ denotes the set of K\% candidate users with a higher score predicted by the audience expansion model w.r.t. the certain campaign. In this paper, following~\cite{liu2020two}, we set $\text{K}=5$.


\vspace{-0.2cm}
\subsection{Offline Results (RQ1)}
In this section, we conduct offline experiments on the two offline look-alike datasets. By treating the actual audiences as positive examples while other candidate users as negative examples, we can evaluate AUC, P@K\%, and R@K\%. From the offline results shown in Table~\ref{tab:offline}, we have several findings.

The two-stage methods which pre-train embeddings outperform the one-stage methods (LR, MLP\_one-stage) which train customized models from scratch. It demonstrates that the embedding pre-training is effective to improve the performance of look-alike modeling. Then, we can find that the third group also outperforms the second group, which validates that pre-training networks are also important.

Another observation is that the gaps between different methods in the  $\mathcal{S}_{[c]} \leq T$ scenarios are larger than in the $\mathcal{S}_{[c]} > T$ scenarios. For the $\mathcal{S}_{[c]} > T$ scenarios, each test campaign has sufficient seed users, and the model can learn relatively sufficient information from the seed set. Thus, the transferable knowledge from other campaigns can only slightly improve the performance. In contrast, for the $\mathcal{S}_{[c]} \leq T$ scenarios, each campaign only has limited seeds, and it is hard to learn a satisfying customized model with the limited seeds. So the transferable knowledge can largely improve the performance of these campaigns.

\begin{table}[t]
  \centering
  \caption{Online A/B testing results.}
%   \vspace{-0.2cm}
    \begin{tabular}{cccc}
    \toprule
    Scenarios & Exposure & Conversion & CVR \\
    \midrule
    video & +3.07\% & +10.18\% & +7.90\% \\
    advertisements & +0.65\% & +15.50\% & +15.40\% \\
    article & +3.18\% & +9.23\% & +4.64\% \\
    \bottomrule
    \end{tabular}%
  \label{tab:online}%
%   \vspace{-0.2cm}
\end{table}%

With the results of the t-test, we can find that MetaHeac could outperform the best baseline significantly in all tasks, which demonstrates that MetaHeac is an effective solution to expand audiences. The main reason is that MetaHeac not only learns how to expand audiences but also takes relationships among various tasks into consideration. MetaHeac outperforms the multi-task methods (Shared-Bottom and MMoE), which demonstrates that the way to learn the relationships among various tasks for the audience expansion problem should be different from traditional multi-task learning. Comparing with the SOTA look-alike methods (Pinterest and Hubble), MetaHeac achieves better results, which validates that pre-training a general model is important for learning customized models for each campaign.

\begin{table}[t]
  \centering
  \caption{Ablation Study on Tencent Look-alike Dataset. HC denotes Hybrid Critics, and HE denotes Hybrid Experts.}
  %\vspace{-0.2cm}
    \begin{tabular}{c||cc||cc}
    \toprule
    \multirow{2}[1]{*}{Method} & \multicolumn{2}{c||}{$\mathcal{S}_{[c]} \leq T$} & \multicolumn{2}{c}{$\mathcal{S}_{[c]} > T$} \\
\cline{2-5}          & AUC   & P@5\% & AUC   & P@5\% \\
    \cline{1-5}
    MetaHeac w/o HC & 0.7199  & 0.2472  & 0.7115  & 0.2220  \\
    MetaHeac w/o HE & 0.7181  & 0.2419  & 0.7112  & 0.2193  \\
    MetaHeac w/o Meta  & 0.7173  & 0.2431  & 0.7107  & 0.2180  \\
    MetaHeac & \textbf{0.7239 } & \textbf{0.2489 } & \textbf{0.7142 } & \textbf{0.2244 } \\
    \bottomrule
    \end{tabular}%
  \label{tab:ablation}%
\end{table}%

\begin{figure*}[!th]
	\centering
	\subfigure[Training Loss]{
		\begin{minipage}[b]{0.23\linewidth}
			\centering
			\includegraphics[width=.95\columnwidth,height=.95\columnwidth]{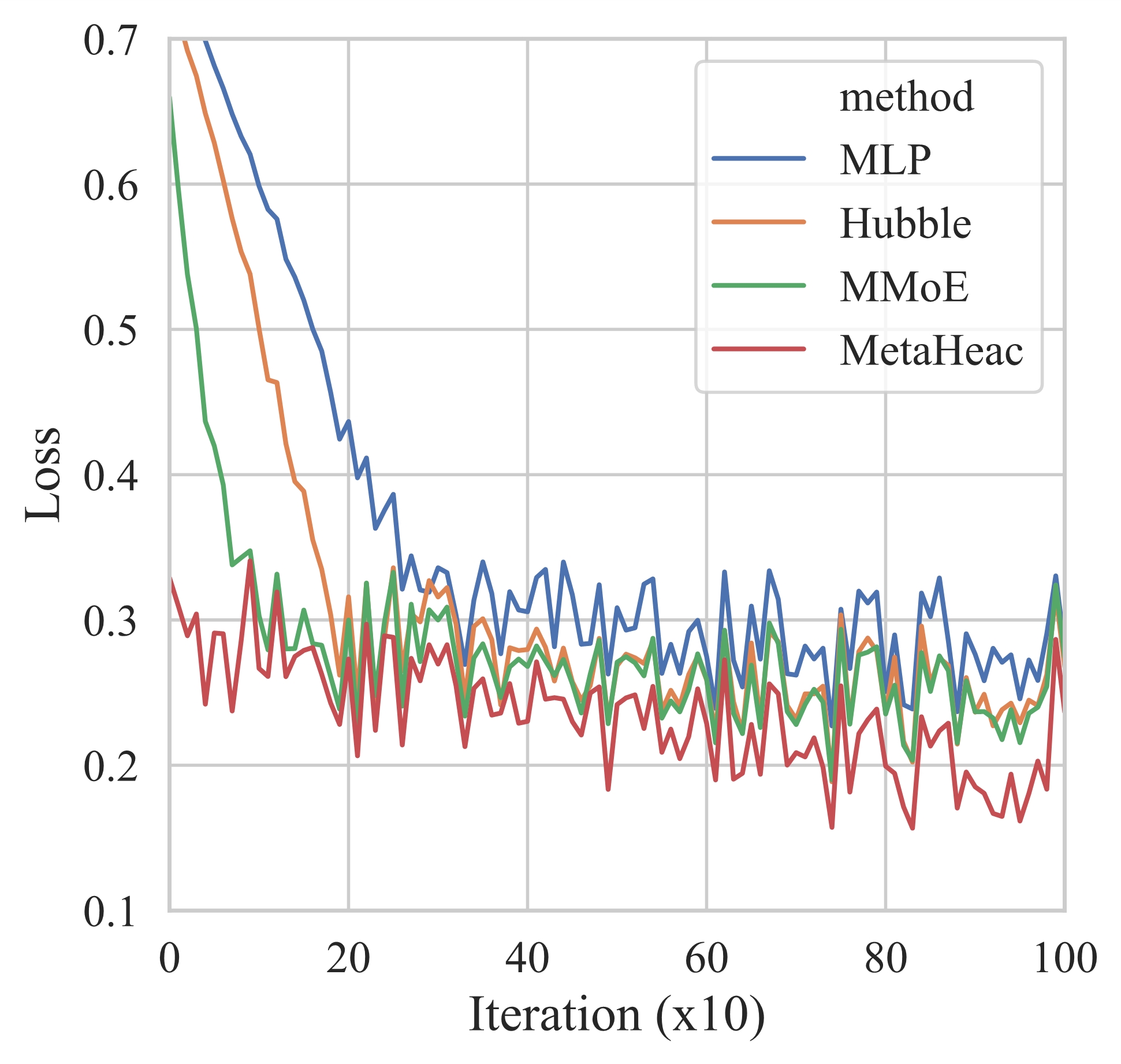}
			\label{fig:3a}
		\end{minipage}
	}
	\subfigure[AUC]{
		\begin{minipage}[b]{0.23\linewidth}
			\centering
			\includegraphics[width=.95\columnwidth,height=.95\columnwidth]{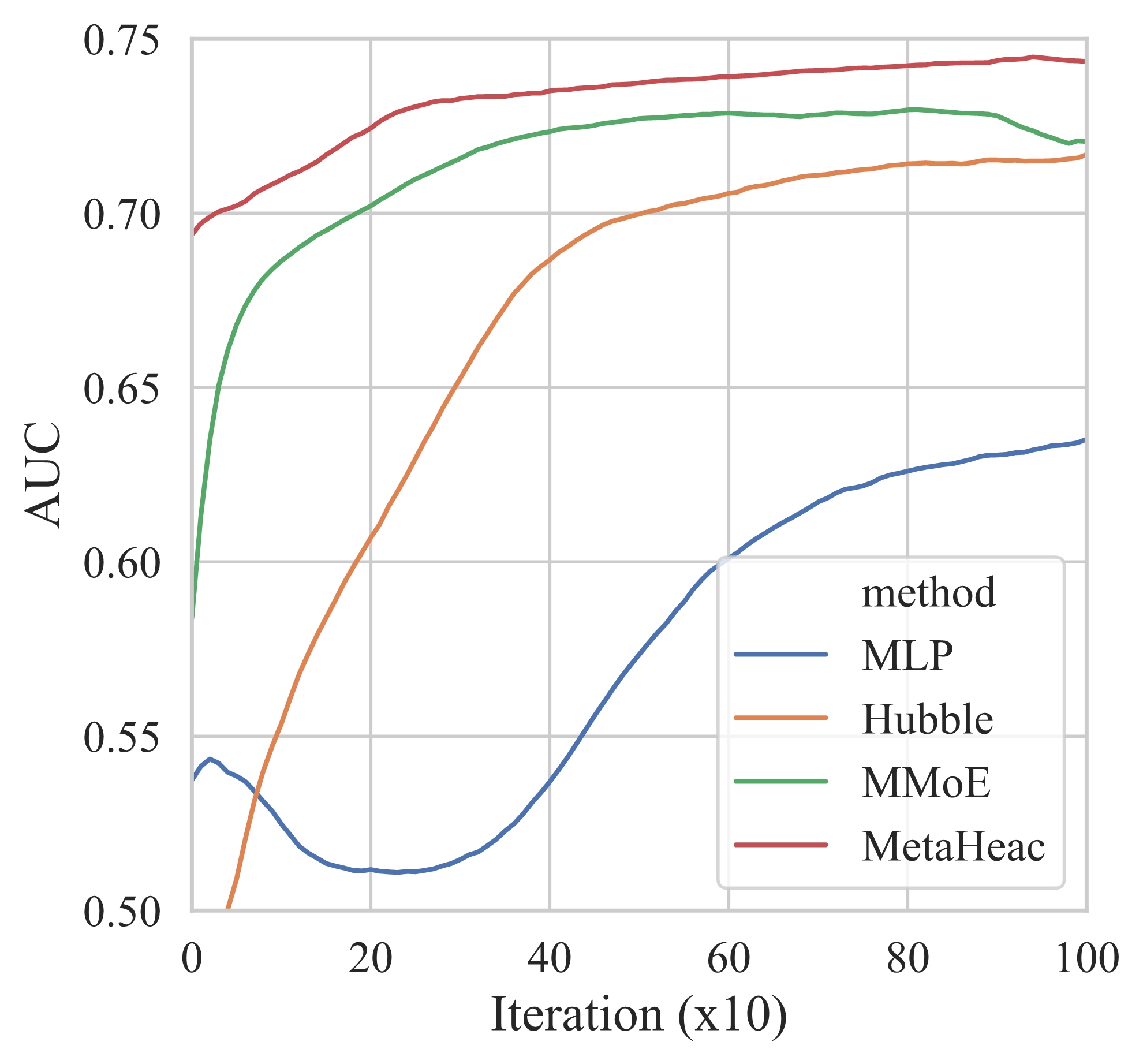}
			\label{fig:3b}
		\end{minipage}
	}
	\subfigure[Average Representations of Seeds]{
		\begin{minipage}[b]{0.23\linewidth}
			\centering
			\includegraphics[width=.95\columnwidth,height=.95\columnwidth]{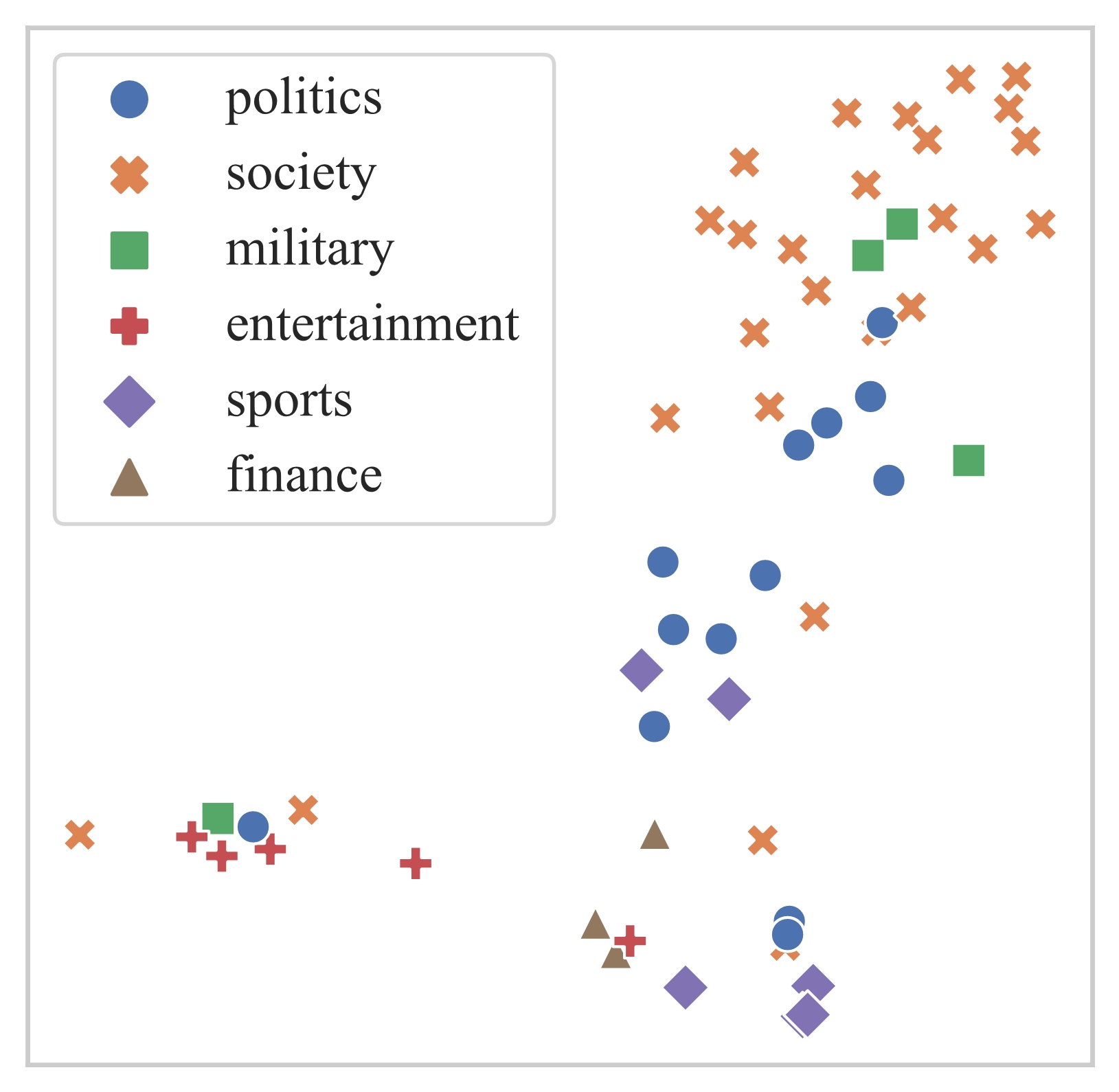}
			\label{fig:3c}
		\end{minipage}
	}
	\subfigure[Gate of Hybrid Critics]{
		\begin{minipage}[b]{0.23\linewidth}
			\centering
			\includegraphics[width=.95\columnwidth,height=.95\columnwidth]{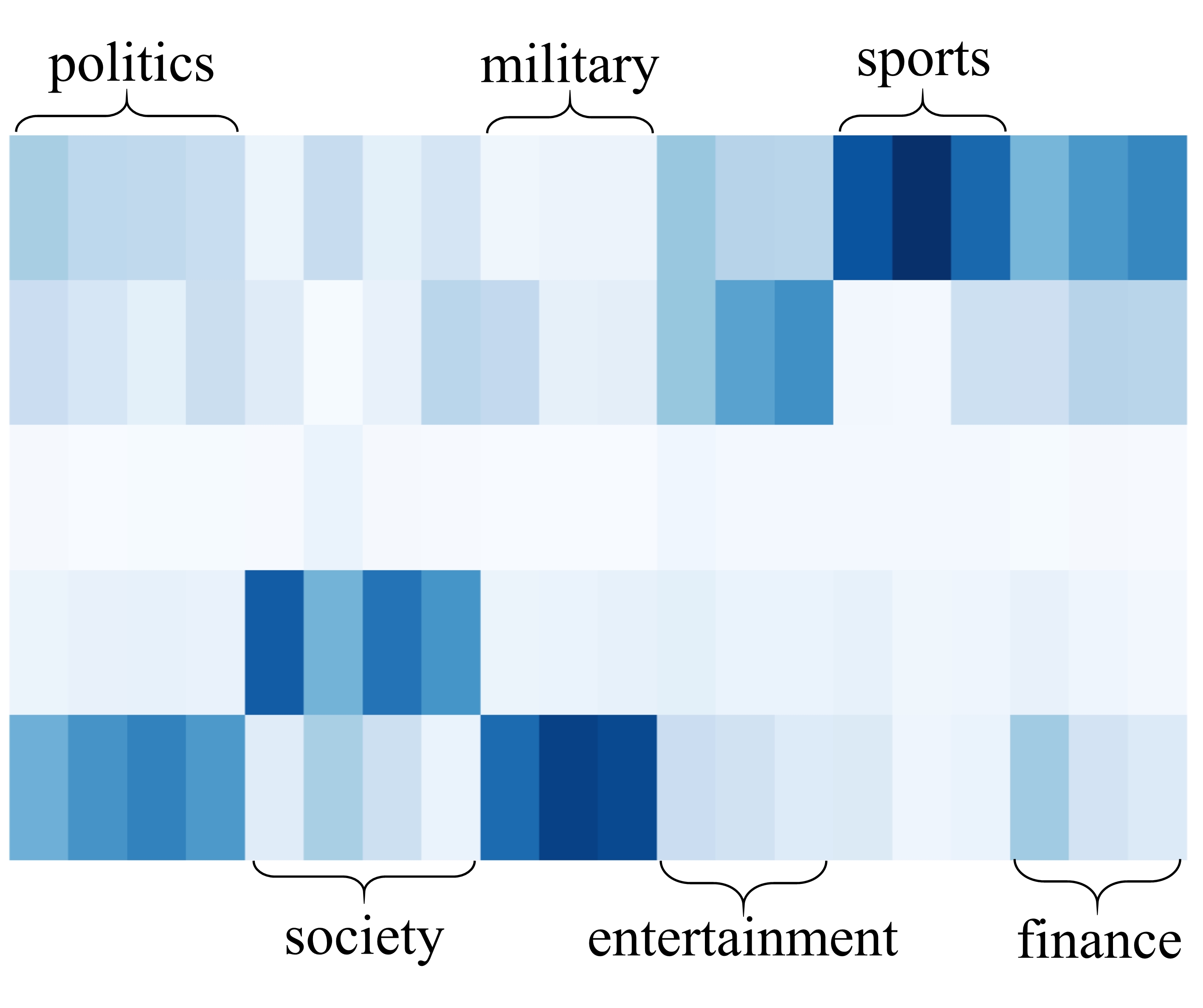}
			\label{fig:3d}
		\end{minipage}
	}
	%\vspace{-0.3cm}
	\caption{(a) Traning Loss and (b) AUC w.r.t. the number of iterations on WeChat Look-alike Dataset. The Visualization of (c) average representations extracted by hybrid experts, and (d) Gate of Hybrid Critics.}\label{fig:analysis}%\vspace{-0.1cm}
\end{figure*}

\subsection{Online A/B Testing (RQ2)}

The MetaHeac framework has already been deployed in WeChat Look-alike System to serve a billion users and conduct hundreds of campaigns. Three applications have utilized the MetaHeac framework to expand audiences, including promotion of video, advertisement, and article in WeChat. 'video' and 'article' are content recommender systems that can conduct content marketing campaigns to promote the specific contents, while 'advertisement' conducts advertising campaigns to promote advertisements. To verify the real benefits MetaHeac brings to our system, we conduct online A/B testing experiments for the three applications within a week. For each application, we randomly split all candidates into two buckets with the same size. The control group here is an MLP\_one-stage classifier as mentioned in subsection~\ref{sec:exp}, while the treatment group is the proposed MetaHeac framework. We allow them to select the same amount of top-scored users from their buckets, respectively. To evaluate the performance online, the following metrics are involved: 1) Exposure: the number of users who have been exposed by the campaign, 2) Conversion: the amount of users who have converted in the campaign, 3) CVR: the conversion rate of the campaign. For the content marketing campaigns, CVR denotes the rate of sharing. The audience expansion task aims to find users who are most likely to finish the business goal of the target campaign. Thus, a higher CVR indicates a higher quality of selected audiences. From the online results shown in Table~\ref{tab:online}, we can see the proposed MetaHeac Framework achieves a significant improvement on Exposure, Conversion, and CVR, against the MLP+pre-training classifier. Especially, the proposed MetaHeac can increase 7.90\%, 15.40\% and 4.64\% CVR for the three scenarios, respectively, which is a very significant improvement and will result in great economic benefit to the company. Thus, our MetaHeac is effective to improve not only content marketing campaigns but also advertising campaigns.

%\vspace{-0.2cm}
\subsection{Analysis (RQ3)}
In this section, we analyze the effects of meta-learning, hybrid experts, and hybrid critics in our proposed MetaHeac. First, we conduct an ablation study to verify the effectiveness of each module. Second, we present convergence analysis to testify the meta-learning framework can converge better and achieve better performance. Third, we visualize the average representations of seeds and the gate of the hybrid critics to show that the hybrid experts and critics can capture the relationships among different tasks.

\textbf{Ablation Study.} To test the effectiveness of each module in MetaHeac, we present an ablation study. Namely the Hybrid Critics introduced in Section~\ref{sec:hc} (HC), Hybrid Experts introduced in Section~\ref{sec:he} (HE) and the Meta-learning Framework introduced in Section~\ref{sec:meta} (Meta). First, we replace the hybrid critics (HC) with a single critic (classifier), which is denoted as MetaHeac w/o HC. Second, we relace the hybrid experts (HE) with a single expert denoted as MetaHeac w/o HE. Third, we replace the training framework of meta-learning with standard BP, which is denoted as MetaHeac w/o Meta. The results are shown in Table~\ref{tab:ablation}, and we have several findings: 1) MetaHeac outperforms MetaHeac w/o HC, which demonstrates that the hybrid critics are effective to improve the performance. 2)  MetaHeac outperforms MetaHeac w/o HE, demonstrating that the hybrid experts are useful to capture multiple representations of users. 3) MetaHeac achieves better results than MetaHeac w/o Meta, which demonstrates training the general look-alike model from the meta-learning perspective could learn how to expand audiences. %(Actually, we also evaluate the results on R@5\%, and we obtain similar conclusions as AUC and P@5\%. Due to the space limitation, Table~\ref{tab:ablation} only contains two metrics.)

\textbf{Convergence.} We testify the convergence of MLP\_one-stage (denoted as MLP), Hubble, MMoE, and MetaHeac, with the training loss and AUC on a random test task in WeChat Look-alike Dataset. We randomly select a task from the test set, and fine-tune the pre-trained model with its training data of seeds and non-seeds (Hubble, MMoE, MetaHeac) or train a new model from scratch (MLP). The training loss and AUC are shown in Figure~\ref{fig:analysis}(a) and (b), respectively.

From Figure~\ref{fig:analysis}(a), we can find that, among these Look-alike methods, MetaHeac achieves the lowest training loss, which testifies MetaHeac has better convergence. Besides, note that MetaHeac has a smaller initial loss, which demonstrates that the meta-learning framework can learn a better initialization. From Figure~\ref{fig:analysis}(b), compared with Hubble and MMoE which learn pre-trained models, MetaHeac achieves a better performance, which demonstrates that MetaHeac can learn more general knowledge from existing tasks. We can find MLP has an extremely unsatisfying performance, and the reason could be overfitting on the training data. The above findings provide empirical evidence that our MetaHeac can learn more general knowledge (how to expand audience), and achieve better convergence and performance.

\textbf{Visualization.} To understand how hybrid experts and critics capture the relationships among various tasks, we visualize average representations of seeds and the distribution of the softmax gate of each campaign task. First, with hybrid experts, we extract seed users' representations and compute the average representation for each campaign task, and then we employ the t-SNE~\cite{donahue2014decaf} to visualize the average users' representations of all tasks, as shown in Figure~\ref{fig:analysis}(c). We can find average representations of the campaigns within the same category are roughly clustering. Note that the campaigns within different categories have no classification boundary. For example, the contents about NBA belong to the category of sports, but it is also related to entertainment. Thus, the campaigns within different categories are also overlapping. We show the average distribution of the softmax hybrid critics' gate of some tasks in Figure~\ref{fig:analysis}(d). Each row denotes the gating weight of a critic, and each column denotes a campaign task. It is easy to observe that the campaign tasks within the same category have a similar distribution of gate. In summary, the above findings demonstrate that the hybrid experts and critics can capture the relationships among various tasks.

%\vspace{-0.5cm}
\section{Related Work}%\vspace{-0.1cm}
In this section, we will introduce the related work from three aspects: Audience Expansion, Meta Learning, Multi-task Learning.

\textbf{Audience Expansion:} Audience expansion (Look-alike modeling) which aims to find similar audiences has been used by many companies. Look-alike approaches can be categorized into three clusters: rule-based, similarity-based, and model-based.

The rule-based methods~\cite{mangalampalli2011feature,shen2015effective} match similar users with specific demographic tags (age, gender, geography) or interests that are targeted by marketers. The core technical support in the background is user profile mining, which means that the interest tags are inferred from the user behaviour~\cite{pandey2011learning}. However, these methods only capture the high-level feature, which is hard to achieve satisfying performance.
%Furthermore, Mangalampalli et al.~\cite{mangalampalli2011feature} built a rule-based associative classifier for campaigns with less conversion. However, these methods only capture the high-level feature, which is hard to achieve satisfying performance.

The similarity-based methods~\cite{liu2016audience,ma2016sub,ma2016score,doan2019adversarial} expand a given seed-set via calculating the similarity of all pairs between seed users and candidate users. The similarity function is usually pre-defined objectively, e.g., Cosine, Jaccard, vector-based dot product. The disadvantage of these methods is that the performance largely depends on artificially defined similarity functions.

%For better computational efficiency, ~\cite{doan2019adversarial} proposed an adversarial factorization auto-encoder that generates binary user representations and encodes higher-order feature interactions. Based on LSH which is often applied to decrease the computation complexity of pairwise similarity~\cite{rajaraman2011mining}, some improved similarity calculation methods~\cite{ma2016score,dewet2019finding} are proposed. The disadvantage of these methods is that the performance largely depends on artificially defined similarity functions.

The model-based methods~\cite{jiang2019comprehensive,qu2014systems} train customized prediction models for each campaign. Our method also falls into this stream. Qu et al.~\cite{qu2014systems} utilized LR to expand audiences, which is effective. In terms of audience expansion performance, ~\cite{jiang2019comprehensive} studied the impact of sampling ratios and sampling techniques. These one-stage methods train customized models from scratch for each campaign which is time-consuming and suffers from the overfitting problem. Recently, some two-stage approaches~\cite{dewet2019finding,zhuang2020hubble,liu2020two} have been proposed to pre-train the embedding with data of all campaigns. However, these methods ignore the generalization ability, task relationships, and deep networks. In contrast, MetaHeac can solve these issues better.

\textbf{Meta Learning:} Also known as learning to learn, meta-learning intends to learn the general knowledge across similar learning tasks, so as to rapidly adapt to new tasks~\cite{finn2017model}. The works in this stream can be grouped into three clusters, i.e., metric-based~\cite{snell2017prototypical}, optimization-based~\cite{finn2017model}, and parameter-generating approaches~\cite{munkhdalai2017meta}. Meta-learning leads to an interest in many areas, such as recommendation~\cite{pan2019warm,lu2020meta,zhu2021transfer,zhu2021learning}, natural language processing~\cite{yan2020multi}, and computer vision~\cite{soh2020meta}. In meta-learning literature, the audience expansion problem relates to general model~\cite{li2018learning} and fast adaptation~\cite{finn2017model}. Since audience expansion is a very practical problem that involves how to update an online system, none of these methods can be directly applied to the online look-alike system.

\textbf{Multi-task Learning:} Multi-task models can jointly learn from several tasks so that it can result in improved efficiency for each task~\cite{ma2018modeling,qin2020multitask,xi2021modeling}. Moreover, multi-task learning is a promising method to learn relationships among different tasks~\cite{ma2018modeling,zhao2019multiple}. However, existing multi-task approaches usually serve for scenarios with fewer than five tasks. In practice, the audience expansion problem faces hundreds of marketing campaign tasks every day, so these multi-task methods cannot work well for this problem.
\vspace{-0.1cm}\section{Conslusion}\vspace{-0.1cm}
In this paper, to solve the audience expansion problem in recommender systems and advertising platforms, we proposed a novel two-stage framework named Meta Hybrid Experts and Critics (MetaHeac). 
%To learn how to expand audiences, we designed a two-phase simulation to train a general model from a meta-learning perspective. To capture relationships among various tasks, MetaHeac employs hybrid experts and critics. 
In the offline stage, a general model which can capture the relationships among various tasks is trained from a meta-learning perspective with existing campaigns. In the online stage, by fine-tuning the general model with given seeds, it is easy for a certain campaign to obtain a customized model to find potential audiences who are likely to finish the business goal.  We demonstrated the superior effectiveness of the proposed MetaHeac in both offline experiments and online A/B testing, compared to other state-of-the-art approaches. Moreover, the MetaHeac framework has been deployed in WeChat for the promotion of both contents and advertisements, which has a significant improvement.
%\vspace{-0.1cm}
\begin{acks}
The research work is supported by the National Key Research and Development Program of China under Grant No. 2017YFC0820604, the National Natural Science Foundation of China under Grant No. 61773361, U1836206, U1811461.
\end{acks}
%\vspace{-0.1cm}

\bibliographystyle{ACM-Reference-Format}

%\input{content/Appendix}

\end{document}